\documentclass[10pt,aps,prd,nofootinbib,reprint]{revtex4-1}

\usepackage[utf8]{inputenc}
\usepackage{amsmath}
\usepackage{graphicx}
\usepackage[colorlinks,pdfusetitle]{hyperref}

\newcommand{\orcid}[1]{\href{https://orcid.org/#1}{#1}}

\newcommand{\e}[1]{\times10^{#1}}

\newcommand{\vev}[1]{\langle {#1} \rangle}
\newcommand{\lsim}{\lesssim}
\newcommand{\gsim}{\gtrsim}
\newcommand{\MPl}{M_{\rm Pl}}

\newcommand{\eq}[1]{Eq.~(\ref{#1})}

\newcommand{\ord}[1]{\mathcal{O}{(#1)}}

\begin{document}

\title{Ultralight Fermionic Dark Matter}

\author{Hooman Davoudiasl}
\email{hooman@bnl.gov}
\thanks{\orcid{0000-0003-3484-911X}}
\affiliation{High Energy Theory Group, Physics Department, Brookhaven National Laboratory, Upton, NY 11973, USA}

\author{Peter B.~Denton}
\email{pdenton@bnl.gov}
\thanks{\orcid{0000-0002-5209-872X}}
\affiliation{High Energy Theory Group, Physics Department, Brookhaven National Laboratory, Upton, NY 11973, USA}

\author{David A.~McGady}
\email{david.mcgady@su.se}\email{dmcgady@alumni.princeton.edu}
\affiliation{Nordita, KTH Royal Institute of Technology and Stockholm University, \\
Roslagstullsbacken 23, SE-106 91 Stockholm, Sweden}

\date{August 14, 2020}

\begin{abstract}
Conventional lore from Tremaine and Gunn excludes fermionic dark matter lighter than a few hundred eV, based on the Pauli exclusion principle.
We highlight a simple way of evading this bound with a large number of species that leads to numerous non-trivial consequences.
In this scenario there are many distinct species of fermions with quasi-degenerate masses and no couplings to the standard model.
Nonetheless, gravitational interactions lead to constraints from measurements at the LHC, of cosmic rays, of supernovae, and of black hole spins and lifetimes.
We find that the LHC constrains the number of distinct species, bosons
or fermions lighter than $\sim 500$~GeV, to be $N \lesssim 10^{62}$.
This, in particular, implies that roughly degenerate fermionic dark matter must be heavier than $\sim 10^{-14}$~eV, which thus relaxes the Tremaine-Gunn bound by $\sim 16$ orders of magnitude.
Slightly weaker constraints applying to masses up to $\sim100$ TeV exist from cosmic ray measurements while various constraints on masses $\lesssim10^{-10}$ eV apply from black hole observations.
We consider a variety of phenomenological bounds on the number of species of particles.
Finally, we note that there exist theoretical considerations regarding quantum gravity which could impose more severe constraints that may limit the number of physical states to $N\lesssim 10^{32}$. 
\end{abstract}

\maketitle

\section{Introduction}
While the astrophysical evidence for dark matter (DM) is overwhelming, its particle nature has evaded detection.
It is generally held that DM cannot be lighter than $\sim10^{-22}$ eV below which point the de Broglie wavelength of DM becomes larger than observed DM structures. On the other end DM heavier than $\sim10^{67}$ eV ($\sim10$ $M_\odot$) would cause significant tidal effects on visible structures. This represents the broadest allowed range of DM mass \cite{Tanabashi:2018oca}.

In 1979, Tremaine and Gunn (TG) pointed out that fermionic DM lighter than $\sim100$ eV would not be contained within a galactic halo \cite{Tremaine:1979we}, immediately ruling out about 24 orders of magnitude of parameter space for fermions.
Modern treatments looking at dwarf spheroidal galaxies find similar bounds on fermionic DM $m\gtrsim50-190$ eV \cite{DiPaolo:2017geq,Savchenko:2019qnn,Pal:2019tqq}.
In this letter, we note that this bound can be significantly weakened if there are many, $N_F$, distinct species of fermionic dark matter whose masses are nearly degenerate.
For very large number of species the momentum of the matter stored within any one of the fermionic species will not exceed the escape velocity from galactic structures and the mass scale of the fermions can be brought down to well below $\cal{O}(\text{eV})$, often considered the \emph{ultralight} regime.  We will thus refer to this possibility as ultralight fermionic dark matter (U$\ell$FDM).  Note that fermion masses can be protected by chiral symmetries and U$\ell$FDM does not pose a fine-tuning problem. 

In what follows, we will first provide an examination of how one could, in a phenomenological fashion, achieve viable U$\ell$FDM and avoid the TG bound in section \ref{sec:TG}.
In section \ref{sec:constraints} we examine some of the potential ramifications of this scenario.  
Importantly and strikingly, we find that the multitude of dark fermions present in our proposal could lead to emergence of deviations in high energy collisions at the LHC, mediated by {\it 4-dimensional} gravity.
Indeed, as emphasized in the literature in recent years, see for example Ref.~\cite{Dvali:2007hz} for interesting early discussions, it seems that some of the strongest bounds on the number of species come from the gravitational equivalence principle.

Additionally, the availability of an enormous number of light fermionic species accelerates evaporation of astrophysical black holes (BHs) and could potentially shorten their lifetimes to astronomical time scales.  Hence, a possible signal of U$\ell$FDM is the detection of sub-solar-mass BHs that are {\it not} of primordial origin.  The above features also yield constraints on the number of U$\ell$FDM species.   A number of the signals that we examine in this work are relevant to ultralight and/or ultra-numerous bosonic, as well as fermionic, species. We will present some of the  corresponding constraints for both.
We then include a broader discussion of scenarios with a large number of species in section \ref{sec:discussion} and then conclude in section \ref{sec:conclusions}.

In appendix \ref{sec:GN} we discuss the relationship between Newton's constant and a large number of species.
We also consider some possibly model-dependent signals of U$\ell$FDM, such as deviations from standard neutrino oscillations in appendix \ref{sec:neutrino} and destabilization of the proton in appendix \ref{sec:proton decay}, at observable levels.
We include a brief review of superradiance physics in appendix \ref{sec:superradiance basics}.
We also provide a proof-of-principle model for production of U$\ell$FDM in appendix \ref{sec:model} and discuss some of its consequences, in particular its implications for structure formation.

\section{Evading the Tremaine-Gunn Bound}
\label{sec:TG}
The TG bound \cite{Tremaine:1979we} is based on the fact that as more fermions are packed into a volume of space, they must span an increasing range of momenta. 
Then, given the escape velocity of a galaxy, this limits the amount of mass that can be squeezed into a galaxy and thus provides a lower limit $m_1\gtrsim100$ eV on the mass of fermionic DM.

\begin{figure}
\centering
\includegraphics[width=\columnwidth]{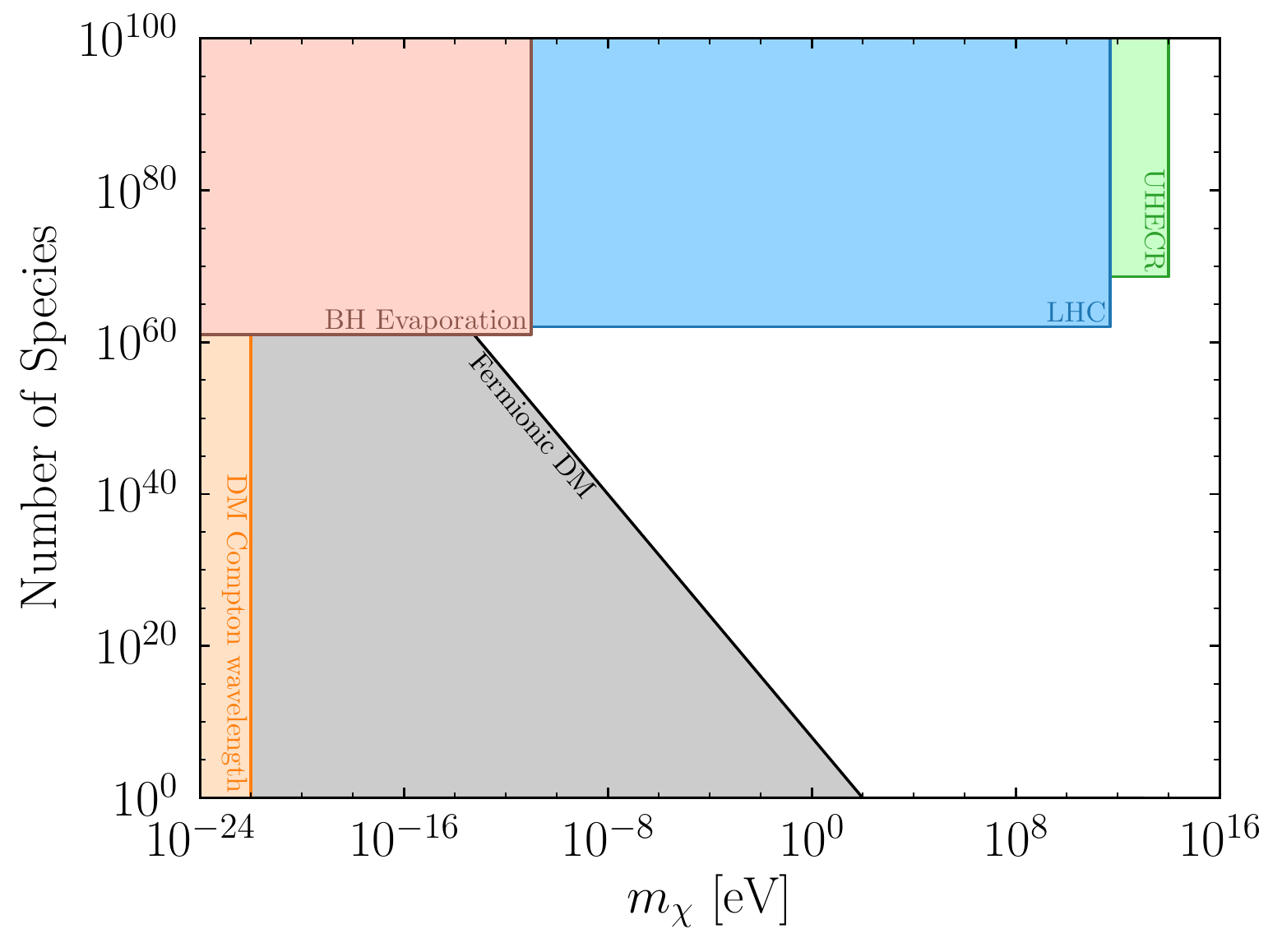}
\caption{A summary of the most stringent constraints relevant for fermionic DM assuming a large number of species with no SM interactions.
For the black shaded region, the constraint only applies for species with quasi-degenerate masses.
Shaded regions are ruled out.
Fermionic DM must have $m\gtrsim10^{-13}$ eV.
Note that we have not included (here or in Fig.~\ref{fig:summary general}) the strong gravity bounds at $N\sim10^{32}$ from e.g.~Ref.~\cite{Dvali:2007hz}.}
\label{fig:summary DM}
\end{figure}

If we consider a scenario wherein DM is composed of many degenerate or quasi-degenerate species, then this bound can be arbitrarily relaxed if we are willing to consider a large number $N_F$ of fermionic fields
\begin{equation}
\mathcal L\supset-m\sum_{i=1}^{N_F}\bar\chi_i\chi_i\,.
\end{equation}
The scaling law on the number of species in order to evade the TG bound is given by \cite{Tremaine:1979we}
\begin{equation}
N_F\gtrsim
\left(\frac{m_1}{m}\right)^4\,,
\label{eq:Ns scaling}
\end{equation}
where $m_1\sim100-200$ eV is the observationally derived constraint on fermionic DM under the assumption of a single species.
Thus $m \lsim 1$ eV requires having $N_F\gsim 10^9$ species.
One of the powers in Eq.~\eqref{eq:Ns scaling} signifies that as DM gets lighter, more of it is required to explain the amount of observed DM while the other three powers come from the phase space.
In principle, this means that the lower bound on fermionic DM could be the same as bosonic DM, $m\gtrsim10^{-22}$ eV, although fermionic DM at this bound requires approaching a googol species, $N_F\gtrsim 10^{96}$.
This constraint is graphically shown in Fig.~\ref{fig:summary DM} along with other strongest constraints relevant for DM.

For typical dwarf spheroidal galaxy masses, the constraint in Eq.~\eqref{eq:Ns scaling} saturates somewhat coincidentally for $m\lesssim10^{-22}$ eV with $N_F\sim10^{96}$ at which point there is one particle from each of the various species and the scaling law  becomes linear in $m$. That is, for $m\lesssim10^{-22}$ eV the $N_F$ scaling is $m^{-1}$.

While U$\ell$FDM can be compatible with observations of galaxies, one must be careful about early universe constraints and ensuring that the DM is cold enough today to be compatible with structure formation.
In appendix \ref{sec:model}, we discuss one way of realizing this scenario.

\section{Constraints on Many Species}
\label{sec:constraints}
Adding a large number of ultralight species opens the door to many interesting constraints. The most general and robust constraints generally derive from gravitational considerations, given their universal coupling to all sources of energy.

\subsection{Gravitational Production}
If there are many light species $N$ (compared with the $\sim100$ species that we are familiar with) that can be produced on-shell at the LHC, {\it i.e.}~with $m \lesssim \ord{1 \rm {\ TeV}}$, then $s$-channel \emph{graviton}-mediated scattering processes can be significant.
By the Equivalence Principle, these graviton amplitudes are blind to whether these states are DM or not.
Within the on-shell S-matrix, the equivalence principle emerges as a consequence of a massless spin-2 particle \cite{Weinberg:1964ew} and allows a one-line computation of the tree-level S-matrix \cite{Benincasa:2007xk,Schuster:2008nh,McGady:2013sga,McGady:2014lqa}.
The gravitational production (tree) amplitude
for the $i^{th}$ species and the total cross section will generally scale as
\begin{equation}
\mathcal A_i\sim \frac{E^2}{\MPl^2}\,,\qquad
\sum_{i=1}^{N} \sigma_i\sim N\frac{E^2}{\MPl^4}\,,
\label{eq:graviton production}
\end{equation}
respectively.  Here, we do not make a distinction between fermions and bosons, since the order of magnitude estimates are not affected at the level of this analysis.  Even though each exclusive cross section is quite negligible at TeV scale colliders, the total cross $\sum_{i=1}^{N} \sigma_i$ can be measurable for sufficiently large $N$. As noted in~\cite{Dvali:2008fd}, this is akin to Kaluza Klein graviton-production in large extra dimensions~\cite{ArkaniHamed:1998rs,Antoniadis:1998ig}.

\subsubsection{Strong Gravity at a TeV}

There are compelling theoretical arguments \cite{Adler:1980bx,ArkaniHamed:1998nn,Antoniadis:1998ig,Dvali:2008fd,Calmet:2008tn,Dvali:2009ne,delRio:2018vrj} indicating that very large numbers of species of quantum fields, $N\gg1$, can lower the energy scale where quantum gravity becomes strongly coupled, well below the Planck mass, $\MPl\approx10^{19}$ GeV.
These arguments have been advanced based on different theoretical principles and could lead to different conclusions about a sensible maximum value for $N$.
The most stringent of these conclusions holds that there are at most $N\sim10^{32}$ states below the TeV scale, regardless of their mass; see for example Refs.~\cite{Calmet:2008tn,delRio:2018vrj}.
This is based on consistency with lack of evidence for strong gravity at the TeV scale, where we have good agreement with the predictions of the SM, which suggests the scale at which gravity becomes strong is $\gsim 1$~TeV.
We will outline some of these ideas in more detail in appendix \ref{sec:GN}.
However, in the rest of this work, we will conservatively concentrate on phenomenological bounds on $N$, based on empirical data, assuming standard general relativity, corresponding to tree-level processes.

\subsection{Cosmology, Colliders, and Supernovae}

\subsubsection{Reheating in Cosmology}

One possible problem in our scenario, akin to that in models with large extra dimensions \cite{ArkaniHamed:1998rs,Antoniadis:1998ig,ArkaniHamed:1998nn}, can occur if the energy density in these $N$ species grows much larger than that of radiation in the early Universe.  As gravitational production in \eq{eq:graviton production} can be substantial, we must ensure it does not populate the Universe with the new states.  For $N$ species, one can show that the ratio of energy densities in the dark states and radiation is given by $N \, T^3/\MPl^3$, where $T$ is the temperature. Since this ratio grows with $T$, it implies a maximum {\it reheat} temperature for any $N$.  To have successful Big Bang Nucleonsynthesis, we require $T\gsim 10$~MeV \cite{Hannestad:2004px}, which yields $N\lsim 10^{63}$.

\subsubsection{LHC}

At the LHC, the production of $\bar \chi_i \chi_i$ via gravitons can be probed if there is initial state radiation.
In that case the experiments would see a monojet plus transverse missing energy $E_{\rm T}^{\rm miss}$. At $E_{\rm T}^{\rm miss}\gtrsim250$ GeV, the total uncertainty on the cross section for invisible processes (such as $\bar\nu\nu$ production via a $Z$), including both experimental and theoretical uncertainties, is $\sim$ 2\% which yields an uncertainty of $\sim 200$~fb, while at $E_{\rm T}^{\rm miss}\gtrsim1$ TeV the combined uncertainty is $\sim$ 10\% or $\sim 1$~fb \cite{Aaboud:2017phn}.  
To estimate the order-of-magnitude for this process' cross-section, we simply divide the parametric $2\to2$ cross section in \eq{eq:graviton production} by $(4 \pi)^3$, which models the phase space suppression for this $2\to 3$ process. We leave more detailed analyses to future work.  

This analysis leads to a constraint $N\lesssim 3\times 10^{65}$ for $m\lesssim125$ GeV.
Via the $E^2$ enhancement in Eq.~\eqref{eq:graviton production}, if $m\lesssim 500$~GeV then the constraint is slightly stronger, $N\lesssim 10^{62}$ \footnote{A similar bound from the LHC was found in Ref.~\cite{Alexeyev:2017scq} which used different methods.}.
This LHC constraint for fermionic DM is presented in Fig.~\ref{fig:summary DM} which requires that fermionic DM must be heavier than $\sim3\e{-14}$ eV.
Further precision measurements at the LHC, high-luminosity LHC, and possible future higher energy circular colliders can all provide improvements on these bounds.

\subsubsection{Ultra High Energy Cosmic Rays}

Ultra-high energy cosmic rays (UHECRs) can also place a constraint on gravitational production. 
On general grounds, increasing $\sqrt{s}$-energy increases gravitational cross-sections to any of the $N$ possible species in hidden sectors. Thus, for large-$N$, UHECRs have an increased likelihood for showers that transfer a large portion of their energy into invisible sectors. This would manifest itself as a suppression (and regeneration) of the flux in terms of visible energy.

Both the Pierre Auger Observatory and Telescope Array observe a significant suppression in the flux above $E_{\rm lab}\sim10^{19.5}$ eV \cite{Deligny:2020gzq} that could be due to either the GZK process \cite{Greisen:1966jv,Zatsepin:1966jv} or the maximum energy of sources \cite{AlvesBatista:2019tlv}. 
We estimate that if $\sim 1/2$ of the events were losing a significant fraction of their energy to the invisible sector this could be noticed as a suppression in the spectrum. Given that the $pp$ total cross section at $E_{\rm lab}=10^{19.5}$ eV is $\sim100$ mb \cite{Tanabashi:2018oca}, we find that UHECR measurements constrain the number of species for $m\lesssim100$ TeV to $N\lesssim4\e{68}$.

\subsubsection{Supernovae}

Since most of the energy from a core collapse supernova is carried away by neutrinos, one could place bounds on emission of feebly interacting particles that would escape the star and deplete its energy. The cross section for neutrino emission is roughly given by $\sigma_\nu \sim E^2 G_F^2$, where $E$ is the typical energy of stellar particles and $G_F \approx 10^{-5}$~GeV$^{-2}$ is Fermi's constant.  The gravitational production of $\chi_i$ is given by $\sigma_\chi \sim N E^2 G^2$. Requiring that only a subdominant amount of the core collapse energy is carried away by the $\chi_i$ species, we then find $N \lsim G_F^2 \MPl^4$ and hence $N \lsim 10^{66}$.  Though this is a rough estimate and subject to various 
astrophysical uncertainties, it suggests that the supernova bounds on $N$ are likely less stringent, for $m\lsim 100$~MeV 
(kinematic limit for emission from a supernova), than those from the LHC data.

\subsection{Black Hole Evaporation}

An interesting consequence of introducing a large number of species is a commensurate reduction in the time it would take for an astrophysical BH to lose a significant amount of its mass due to Hawking evaporation \cite{Hawking:1974rv,Hawking:1974sw}.  Under the usual assumptions about the number of long wavelength species, basically the photon plus the graviton, one would expect an enormous lifetime for a solar mass black hole given by \cite{Page:1976df}
\begin{equation}
t_{evap} \approx 10^{67} \left(M_{\rm BH}/M_{\odot}\right)^3 \; \text{yr}\,,
\label{tevap}
\end{equation}
where $M_{\rm BH}$ is the BH mass and $M_\odot \approx 2\times 10^{33}$~g is the solar mass~\cite{Page:1976df}.
Up to factors of order unity, adding $N$ more particles will result in $t_{evap}\to t_{evap}/N$~\cite{Dvali:2007hz,Dvali:2008ec,Dvali:2007wp,Holdt-Sorensen:2019tne}.  We see that for the number of particles allowed by the LHC data, $N \sim 10^{62}$, the lifetime of a solar mass black hole will be short on cosmic time scales.  

There is no clear observational lower bound on the lifetime of a BH known to us.  However, the $\sim 10 M_\odot$ BH mergers observed through gravitational waves by LIGO and Virgo collaborations \cite{LIGOScientific:2018mvr} would be very hard to explain if the lifetime of such BHs  was much smaller than $\sim 10^9$~yr.  The above \eq{tevap} then suggests that a plausible bound from the observation of those mergers is $N \lsim 10^{61}$, which is similar to the current other constraints.

Here, we note that BH mass loss due to Hawking evaporation is in principle not feasible under standard assumptions.  This is because the ambient cosmic microwave background radiation (CMBR) represents a much hotter thermal system, with a temperature $T_{\rm CMB}\sim 10^{-4}$~eV, than a solar mass black hole with $T^\odot_{\rm BH}\sim 10^{-11}$~eV.  Hence, under standard assumptions, BHs cannot presently lose mass, due to their immersion in the CMBR.  However, with a large number of species, $N \gg (T_{\rm CMB}/T^\odot_{\rm BH})^4$ typical of the discussion in this work, the rate of particle emission will far surpass that of CMBR photon absorption and the above reduction of $t_{evap}$ can occur in the current cosmological epoch.   

\subsection{Superradiance}
Superradiance is the process wherein light particles are produced on-shell out of the vacuum near a rotating BH extracting angular momentum from it forming a cloud \cite{Penrose:1969pc}.
This process only efficiently happens for particle masses on the same length scale as the radius of the BH.
When BHs with large spins are measured, this implies that the BH has not spun down too much and thus particles of the relevant mass do not exist.
Constraints have been derived for bosons using observations of stellar mass and super-massive BHs (SMBHs)  \cite{Baryakhtar:2017ngi,Davoudiasl:2019nlo,Brito:2015oca,Zu:2020whs}.
Due to fermionic degeneracy pressure this process will not happen for a single species of fermion, but can happen for many species,  as discussed below.  For many bosonic species this process is accelerated compared to the single boson scenario.  We conservatively take the constraints on bosons to correspond to spin-0 particles and note that the lower mass limit strengthens as the number of species increases as $m^{-9}$. 

\begin{figure}
\centering
\includegraphics[width=\columnwidth]{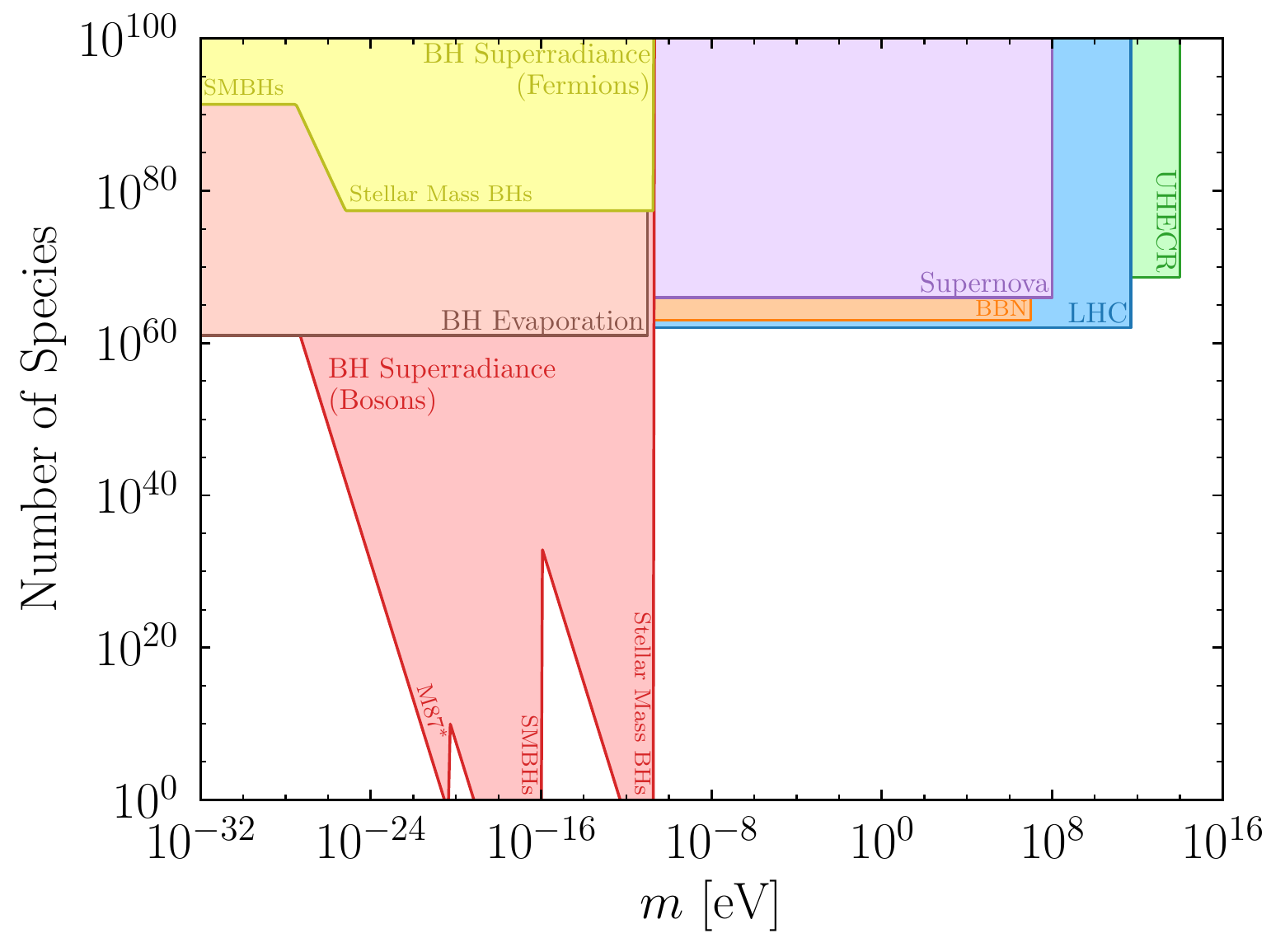}
\caption{A summary of several constraints relevant for any new physics model assuming a large number of species with no SM interactions; shaded regions are ruled out. 
All constraints apply regardless of spin, unless  specified.}
\label{fig:summary general}
\end{figure}

For fermions, the constraints on $N_F$, up to modest numerical factors, scale like $m^{-6}$, but only apply when there are more species than the occupation number in the cloud which scales as $N_{\rm occ}\propto M_{\rm BH}^2$.  We provide some of the equations relevant to superradiance phenomena considered here, in appendix \ref{sec:superradiance basics}.  The relevant constraints for many boson and fermion species are plotted in Fig.~\ref{fig:summary general}. 

We conservatively take the scalar limits derived for $N_B=1$ and $m^{-9}$ scaling, for bosons.
For fermions, we take the same limits as at $N_B=1$ and follow the $m^{-6}$ scaling.
For fermions, however, Pauli exclusion pressure dictates that $N_F \sim N_{\rm occ}$.  
From Ref.~\cite{Baryakhtar:2017ngi}, one finds that $N_{occ}\sim10^{77}(M_{\rm BH}/10\,M_\odot)^2$.

For stellar mass BHs we use the $N_B=1$ bounds from Ref.~\cite{Baryakhtar:2017ngi}, for SMBHs we use the bounds from Ref.~\cite{Brito:2015oca}, and for M87$^*$, we use the bounds from Ref.~\cite{Davoudiasl:2019nlo}.
These correspond to the three disjoint regions at $N_B=1$ for bosons in Fig.~\ref{fig:summary general}; for sufficiently large $N_B$ they all overlap.
For fermions, the majority of the region in this figure is due to stellar mass BHs, while for $m\lesssim10^{-27}$ eV the constraint is from SMBHs.
Unlike the other constraints in the figures, the fermionic superradiance constraint does not trivially continue to arbitrarily smaller masses, as they continue to step up in the same way given by the $m^{-6}$ scaling law for the sloped regions, and the occupation number of the cloud for the flat regions.

\section{Discussion}
\label{sec:discussion}
Our main motivation for considering an enormous number of species $N$ in this work has been to accommodate a phenomenologically viable scenario for ultra light fermionic DM.  However, large values of $N$ have also been considered in relation to addressing other theoretical problems, in particular the question of the apparent hierarchy between the weak scale and the scale of new physics, which could presumably extend all the way to $\MPl$ \cite{Dvali:2007hz,Arkani-Hamed:2016rle}.  Such proposals do not necessarily entail a large number of ultralight states. 

For example, the scenario in Ref.~\cite{Dvali:2007hz} would require $N\sim 10^{32}$ states with masses $m\sim 1$~TeV in order to explain the large value of Planck scale, via 
the relation $\MPl^2 \sim N m^2$. See Ref.~\cite{Dvali:2009ne} for further  phenomenological examination of this idea.  An even earlier proposal that entails a large number of new degrees of freedom is that of large extra dimensions, where a large number of light, compared to the weak scale, Kaluza Klein gravitons are present in the 4-dimensional theory \cite{ArkaniHamed:1998rs,Antoniadis:1998ig}.  

In Ref.~\cite{Arkani-Hamed:2016rle}, on the other hand, it is assumed that if a very large number of copies of the SM are available, then an accidentally small value of the weak scale for some of the copies could typically arise.  Here, the copies are non-interacting, except gravitationally; for instance, if they could be localized on separate branes distributed along an extra dimension (see also Ref.~\cite{ArkaniHamed:1999zg}).
Interestingly this model \emph{predicts} a very large number of light $\lesssim100$ eV fermions with masses that may well go down to $\sim10^{-10}$ eV if the neutrino copies have only Dirac mass terms.
The proposal in Refs.~\cite{Dienes:2011ja,Dienes:2011sa} also considers numerous species making up DM, however they do not focus on ultralight fermions; a typical realization of this proposal is based on  
theories with large extra dimensions where the large number of Kaluza-Klein states contribute to the DM population.
Another mechanism for generating a large number of light species is via freeze-in \cite{Cohen:2018cnq}.

Our work shares various elements with the above constructs, but is motivated from a different point of view, and it could in principle be complementary to them.  For example, in our work, $N_F\sim 10^{60}$ light fermions with $m \sim 10^{-13}$~eV could constitute ultralight DM, but this does not necessarily address the hierarchy problem, since $N m^2 \ll \MPl^2$, in the framework presented in Ref.~\cite{Dvali:2007hz}, for example. Nonetheless, our scenario is not exclusive of having a large number of heavier states of mass $\gsim 1$~TeV that could potentially address the hierarchy between the weak and the Planck scales.  

The requisite large values of $N$ may possibly originate from a full theory of quantum gravity, for example from a large number of branes in extra dimensions, alluded to before.  From an effective theory point of view, one could also imagine that 
$N\gg 1$ flavors could appear if the states form representations of $\ord{k}$ distinct 
flavor symmetry groups of $SU(n)$ type.  With $k \sim 10-100$ and $n \sim 10$, one finds $N\sim n^k$; similar ideas have been entertained regarding fermionic DM, for modest values of $N \sim \ord{10}$ \cite{Randall:2016bqw}.  We do not further speculate on the origin of $N$, and suffice it to present a few possibilities.

Scenarios with a large number number of new states have also been  
examined with respect to cosmological evolution \cite{Horvat:2008ze}, in the context of the proposal in Ref.~\cite{Cohen:1998zx} which  addresses the problem of a finely-tuned cosmological constant.

\section{Conclusions}
\label{sec:conclusions}
In this work, we have considered the possibility that, in contrast to the canonical view, dark matter could be composed of ultralight fermions much lighter than the often-assumed Tremaine-Gunn  $\ord{100~\text{eV}}$ lower bound.  That lower bound, based on phase space density of fermions, implicitly assumes that all DM is made of a single species.  If one allows for the dark matter population to comprise a large number $N_F$ of fermions, the phase space restrictions which depend on spin statistics of identical fermions (Pauli's exclusion principle) are avoided and the mass lower bound can be relaxed with $N_F^{-1/4}$.  

Our ultralight fermion dark matter scenario has a number of striking phenomenological consequences.  Due to the enormous number of 
species one could expect gravitational effects that are normally completely negligible.  These include detectable effects from graviton exchange at the TeV scale, such as in collider experiments or via high energy cosmic rays, as well as accelerated evaporation of solar-mass black holes on astronomical time scales.  We considered these and other effects, and found that they roughly yield the constraint $N\lsim 10^{62}$, for which one could accommodate fermionic dark matter as light as $10^{-13}$~eV.  Depending on the assumptions of the underlying model, stronger bounds could apply, for example originating from the possibility of gravity-mediated fast proton decay or non-standard neutrino oscillations, as well as possible modifications of Newton's constant (see the appendices).  

Our work illustrates that departure from a monolithic picture of fermionic dark matter, by allowing a large number of species, could open novel and exciting phenomenological possibilities that deserve attention, as the decades-long search for clues to the identity of dark matter continues.

\begin{acknowledgments}
{\bf Acknowledgements:} HD and PBD are supported by the US Department of Energy under Grant Contract DE-SC0012704. DAM is supported by a Nordita fellowship that is itself supported by the grant “Exact Results in Gauge and String Theories” from the Knut and Alice Wallenberg Foundation.
\end{acknowledgments}

\appendix

\section{Possible Modification of Newton's Constant from Quantum Effects of Many Species}
\label{sec:GN}

For the sake of clarity, we briefly review some possible one-loop
effects on the graviton coupling, {\it i.e.}, Newton's constant, relevant
to the “Strong Gravity” species bounds discussed in the main text.

In Ref.~\cite{Dvali:2007hz}, based on a BH evaporation gedanken experiment, it is argued that $N$ states of mass $\sim m$ would contribute a quantum correction $\sim N \, m^2$ to $G^{-1}$ which is renormalized to yield the low energy value inferred from long distance gravity. This interpretation seems consistent with the discussion in Ref.~\cite{Adler:1980bx}. 
Here, the running of $G(\mu)$, where $\mu$ is the renormalization scale and $\mu > m$, would presumably be logarithmic, schematically of the form 
\begin{equation}
G^{-1}(\mu) \sim G^{-1}- N\, m^2 \,\ln(\mu^2/m^2)\,,
\label{GNlog}
\end{equation}
where $G^{-1}\equiv G^{-1}(0) = \MPl^2$.   

In the limit where all $N$ species are of similar masses $m$, Planck-weak naturalness arguments presented in Ref.~\cite{Dvali:2007hz} would then suggest that 
\begin{equation}
m \lsim \MPl/\sqrt{N}\,.
\label{m-upper}
\end{equation}

A different analysis in~\cite{Calmet:2008tn} yields a different result for this running for $\mu> m$:
\begin{equation}
G^{-1}(\mu) = G^{-1} - N \,\mu^2/12 \pi\,.
\label{GNmu2}
\end{equation}
This expression reproduces the $N \lesssim 10^{32}$ upper-bound discussed previously.  That is, Eq.~\eqref{GNmu2} implies that at $\mu\gsim$~TeV, independent of the mass scale $m$ of the new species, we reach strong quantum gravity for $N\sim 10^{32}$.  

Apart from its high energy manifestations, a running $G$ could also lead to deviations from Newtonian gravity at macroscopic distances. Currently, such deviations are constrained to be $\delta G/G \lsim 10^{-9}$, at distances larger than $\sim 10^3~\text{km}\sim (10^{-13}~\text{eV})^{-1}$ \cite{Fayet:2017pdp,Schlamminger:2007ht}.  The running behavior in \eq{GNmu2}, which implies $N\lsim 10^{32}$ as discussed above, is not further constrained by these measurements.  

In the case of running according to \eq{GNlog}, consistency with current bounds could be achieved for sufficiently small  $\delta G/G \sim N \, m^2\, G$.  For example, if $N\sim 10^{60}$ and $m\sim 10^{-13}$~eV, as possible values suggested by our phenomenological analysis, then $\delta G/G \sim 10^{-22}$, which is negligible and hence not detectable by long distance tests of gravity.  However, if $N\sim 10^{60}$ and $m\sim 10^{-3}$~eV, as favored by a naturally large Planck scale according to \eq{m-upper}, one would roughly have $\delta G/G \sim 10^{-2}$, which is close to the current bounds on deviations from Newtonian gravity at length scales $1/m \sim 0.1$~mm \cite{Lee:2020zjt} and could potentially be tested in the future.  In this case, however, there would be no deviations at larger distances corresponding to $\mu \ll m$, where $G$ does not run.

\section{Neutrino Oscillations}
\label{sec:neutrino}

Here, we give a schematic analysis of how the additional $N_F$ species could alter the physics of neutrino oscillation.  There are a variety of model-dependent possibilities, however here we choose a simple setup as an example.  Let us assume that SM neutrino masses $m_\nu\sim 0.1$~eV are generated through coupling to heavy right-handed Majorana neutrinos $N_R$ that are complete singlets and couple to a left-handed lepton doublet $l$ and the Higgs field $H$  
through a Dirac mass term: $y_N \, H^* \bar l \, N_R$; we have suppressed generational indices and $y_N$ is implicitly a Yukawa matrix.  For concreteness, we will take 
$y_N \sim 1$ and $m_R \sim 10^{14}$~GeV for $N_R$ mass.  

Let us assume that extra $N_F$ fermionic species $\chi_i$ are nearly degenerate with masses $\ord{m} \ll m_\nu$; $\chi_i$ are not necessarily DM, but they could well be.  The following coupling between the new species and the SM is generally allowed,
\begin{equation}
\xi_i \, H^* \bar l \, \chi_i\,.
\label{Diracmass}
\end{equation}
  
Neutrino flavor oscillations $\nu_\ell \to \chi_i$, with $\ell$ denoting a lepton flavor, $e,\mu,\tau$, are then characterized by the probability
\begin{equation}
P(\nu_\ell \to \chi_i) \sim \sin^2 (2 \theta_i)
\sin^2\left(\frac{m_\nu^2 \, L}{4E}\right)\,,
\label{Pnuosc}
\end{equation}
where $L$ here is the baseline length and $E$ is the neutrino energy; the mixing angle $\theta_i$ is given by 
\begin{equation}
\theta_i \sim \frac{\xi_i \vev{H}}{m_\nu}\sim 10^{12} \xi_i\,.
\label{thetai}
\end{equation}

The total probability for disappearance of $\nu_\ell$ into one of the $N_F$ species is then given by 
$P \approx \sum_{i=1}^{N_F} P(\nu_\ell \to \chi_i)\sim N_F P(\nu_\ell \to \chi_i)$; we have assumed $\xi_i\approx\xi$ and hence $\theta_i \approx \theta\, \forall\; i$, for simplicity.  Requiring that the current experimental results do not show large deviations from the standard picture \cite{Parke:2015goa} leads to the generic bound $N_F \theta^2 \lesssim0.1$ unless the lightest active neutrino is very light.  
For example, with $N_F\sim 10^{60}$ - typical of some of our bounds in this work - we find $\xi \lesssim \ord{e^{-97}}$.  Such a small number could imply a suppression from ``instanton" mediation,  for example due to the violation of a global charge  carried by $\chi_i$ through non-perturbative gravitational effects; see {\it e.g.} Refs.~\cite{Abbott:1989jw,Kallosh:1995hi,Svrcek:2006yi} and Ref.~\cite{Davoudiasl:2020opf} for a recent discussion in the context of neutrino mass generation.  With $\xi \sim e^{-S}$, where $S$ is the action for the instanton, one gets $S\gsim 97$.       
The value of $S$ depends on the underlying ultraviolet theory of gravity, but $S\sim 100$ could be typical of such 
theories \cite{Hui:2016ltb}. (Here, it is implicitly assumed that the addition of a large number of new species still allows for the instanton action to have the requisite value, in units of $\hbar$.) 

\section{Proton Decay} 
\label{sec:proton decay}

An interesting aspect of this proposal is that there could potentially be a large enhancement of gravity mediated processes.  In particular, proton decay - typically 
considered unobservable if suppressed by Planck mass - may become accessible in our scenario.  This could, however, depend on whether proton decay operators carry global charges and how such charges are destroyed by gravitational processes, as discussed above.  For example, let us consider the operator 
\begin{equation}
O_n \sim \frac{u d d \chi_i}{\MPl^2}\,,
\label{uddpsi}
\end{equation}
where $u$ and $d$ represent up and down type quarks and $\chi_i$ is one of the fermionic species of ultralight DM.  

The above interaction mediates neutron mixing with $\chi_i$.  Here, it is implicitly assumed that $\chi_i$ are not charged under a gauged symmetry.  We will address the case of global symmetries below.  On general grounds, 
we may expect that $O_n$ can be generated by gravitational interactions.  One could roughly estimate the rate for $p\to \pi^+ \, \bar \chi$ from $O_n$ above, by 
\begin{equation}
\Gamma(p\to \pi^+ \, \bar \chi_i) \sim N_F \frac{m_p^5}{\MPl^4}\,,
\label{pdecay}
\end{equation}
where $m_p\approx 0.94$~GeV is the proton mass.  To avoid a lifetime much shorter than $\ord{10^{32}}$~yr, roughly around the current limits \cite{Tanabashi:2018oca}, we find that $N_F \lsim 10^{12}$, 
which is quite a strong bound and limits fermionic DM masses to scales $\gsim 0.1$~eV, from \eq{eq:Ns scaling}.  Based on kinematic considerations alone, the proton decay bound is relevant up to $\chi_i$ masses below $\sim m_p - m_\pi \sim 800$~MeV, however the above rough estimate assumes no phase space suppression and is hence valid roughly for $\chi$ masses lighter than $\sim 100$~MeV.   

If the operator in \eq{uddpsi} carries a global charge, it could require violation of a $U(1)$ global symmetry by some type of gravitational instanton, as in the preceding discussion (see again Refs.~\cite{Abbott:1989jw,Kallosh:1995hi,Svrcek:2006yi}).  This would suppress the rate for decay by $e^{-2 S}$,  potentially severely suppressing the rate in \eq{pdecay} and making the constraints from proton decay ineffective compared to other considerations.  See also Ref.~\cite{Dvali:2009ne} for a discussion relevant to a scenario that could address a natural Planck-weak hierarchy \cite{Dvali:2007hz} with a large number of species. 

\section{Superradiance Basics}
\label{sec:superradiance basics}

Here, we provide a few expressions at the order-of-magnitude level relevant to the derivation of the superradiance constraints.  For the bosonic case, only one sufficiently light boson speices can lead to exponential growth of its amplitude.  We will consider $N_B$ species of scalars, since they lead to a more conservative bound compared to vector bosons.  For more details of requisite superradiance condition and related physics see, for example, Refs.~\cite{Arvanitaki:2009fg,Baryakhtar:2017ngi}. 
The dominant rate for emission of $N_B$ species of degenerate scalars of mass $m_S$ from a spinning BH is roughly given by
\begin{equation}
\Gamma(S) \sim N_B a\, r_g^8\, m_S^9\,,
\label{GamScalar}
\end{equation}
where $a$ is the dimensionless spin parameter of the BH of mass $M_{\rm BH}$ and $r_g \equiv G\, M_{\rm  BH}$.

The spin of the BH can efficiently change by $\Delta a$, via emission of light states roughly satisfying $m_S \sim 1/r_g$, up to occupation number $N_{\rm occ}$, where 
\begin{equation}
N_{\rm occ} \sim r_g M_{\rm BH} \Delta a\,,
\label{calN}
\end{equation}
subject to the condition
\begin{equation}
\Gamma(S)\, \tau_{\rm BH} \gsim 
\ln N_{\rm occ}\,.
\label{spin-cond}
\end{equation}
In the above relation, $\tau_{BH}$ 
is the ``timescale" of the BH. 
A typical relevant timescale is the Salpeter time $\tau_{\rm BH} \sim 5\times 10^7$ years and is based on the Eddington limit \cite{Salpeter:1964kb,Baryakhtar:2017ngi}.

In the case of $N_F\gg 1$ species of fermions considered in this work, we adapt the expressions for spin-1/2 particles $\chi_i$, following for example Ref.~\cite{Baryakhtar:2017ngi}.  We find (at the order-of-magnitude level),
\begin{equation}
\Gamma(\chi) \sim N_F a\, r_g^5\, m^6\,.
\end{equation}
Note that due to the Pauli exclusion principle, $N_F\gtrsim N_{\rm occ}$ is needed for fermionic particles.

We note that just because a BH could have produced $N_F$ particles, does not mean that they are all likely to be distinct.
If $n$ elements are randomly sampled with replacement from a set size $n$, the average number of distinct elements is given by $(1-1/e)n\approx0.63n$ for large $n$.
Thus, only $\sim37\%$ of the events are likely to be repeats and the combinatoric issue can be safely ignored as its effect is less than a factor of two.

\section{A Model}
\label{sec:model}

We now describe a mechanism for realizing the ultralight fermionic DM scenario described above.  In general terms, we demand that this mechanism lead to the correct energy density of DM by the epoch of matter-radiation equality, characterized by a temperature $T_{eq}\sim 1$ eV.  Furthermore, we will demand that at $T\sim T_{eq}$, the DM population be sufficiently cold which we will take to correspond to velocities $v_\chi\lsim 10^{-3}$.

The main production mechanism assumed here is the decay of an axion whose mass $m_a$ is close to the common mass scale $m$ of the DM species: $m_a \gsim 2 m$.
We assume that the axion couples approximately universally to all $N_F$ species with coupling $g_\chi\sim m/f_a$ and that all $N_F$ species are almost degenerate in mass; here $f_a$ is the axion decay constant\footnote{This coupling could arise in several ways.
Perhaps there is a complex scalar $\Phi$ charged under $U(1)_X$ and the fermions are chiral and are also charged under $U(1)_X$.
If $\Phi$ gets a vev it gives mass to all the fermions.
If the Yukawa couplings are all the same order of magnitude, then the phase of $\Phi$, the axion, would also couple to all the fermions with roughly uniform couplings as well.}.
As the axion oscillates and decays, it populates all $N_F$ species roughly equally.  

Once produced through axion decay, the fermions  will free-stream. Let us denote the temperature corresponding to the axion decay, assumed nearly instantaneous, to be 
$T\sim T_d$.  The fermions are initially produced with some boost, and they will start to be non-relativistic at $T\sim T_{NR}$, given by $T_{NR}\sim (m/m_a) T_d$.  By the time $T\sim T_{eq}$, the fermions will have a typical velocity given by
\begin{equation}
v_\chi(T_{eq}) \sim \frac{T_{eq}}{T_{NR}}\,.
\label{vchi}
\end{equation}
It then suffices to have $T_{NR} \gsim 1$~keV to achieve $v_\chi\lsim 10^{-3}$.  

Note that the co-moving free-streaming length $\lambda_{FS}$ for $\chi_i$ is given by 
\begin{equation}
\lambda_{FS} \approx \int_{t_d}^{t_{eq}} \frac{v_\chi(t)}{a(t)}\, dt\,,
\label{lFS}
\end{equation}
where the lower limit of integration corresponds to the time $t_d \approx 0$ of the axion decay, assumed small compared to other time scales.  The upper limit $t_{eq}$ corresponds to the time of matter-radiation equality; $a(t)$ is the scale factor for the expansion of the Universe.  We have
\begin{equation}
v_\chi(t) \approx \frac{c\, a(t_{NR})}{a(t)}\,,
\label{vt}
\end{equation}
where $c$ denotes the speed of light, made explicit for clarity of presentation; $t_{NR}$ corresponds to $T_{NR}$ when the fermions become non-relativistic. We find 
\begin{equation}
\lambda_{FS} \approx \frac{2\, c\,  t_{NR}}{a(t_{NR})}\left[1 + 
\ln \left(\frac{a(t_{eq})}{a(t_{NR})}\right)\right]\,.
\label{lFSexp}
\end{equation}
For $T_{NR}\gsim 1$~keV, we find $\lambda_{FS}\lsim 300$~kpc.  Since structures on scales $\lambda_{FS}$ and smaller get suppressed by free streaming, the above suggests that for $T_{NR}\gsim 1$~keV structure formation at the galactic size is viable\footnote{Note the bottom-up structure formation, starting from sub-galactic halos, may require $T_{NR}\gsim 10$~keV.}. 

Let us examine, as an example, the case where $N_F\sim 10^{60}$ which is near interesting collider and astrophysical signals, as discussed earlier.  In particular, for $m \sim  10^{-11}$~eV $\sim (20~\text{km})^{-1}$, a possible value for DM mass, astrophysical black holes could emit $\chi_i$ without suppression, leading to their fast evaporation.  The width (inverse lifetime) of the axion from decay into $N_F$ degenerate fermions is given by 
\begin{equation}
\Gamma_a \sim N_F\, \frac{g_\chi^2 \, m_a}{8 \pi}\,.
\label{Gamma-axion}
\end{equation}
We have $\Gamma_a \sim H(T_d)$, with the Hubble scale $H\sim g_*^{1/2} T^2/M_P$, where $g_*$ is the number of relativistic degrees of freedom at temperature $T$.
The initial energy density stored in the axion, assuming a typical initial amplitude $\ord{f_a}$, is $\sim m_a^2 f_a^2$ and redshifts like $T^3$.  

The axion energy will later convert into a ensemble of relativistic fermionic DM states (for $m_a\gsim 2 m$) whose energy density redshifts as $T^4$.  Once the fermions become non-relativistic at $T\sim T_{NR}$, their energy density will again decrease like matter, that is with $T^3$.  Hence, for DM fermions to arrive at the era of matter-radiation equality with the correct energy density, the requisite initial temperature $T_i$ when the axion must start oscillating is given by
\begin{equation}
T_i \sim m_a m \left(\frac{N_F \MPl}{8 \, \pi g_*^{1/2} m_a\, T_{eq} T_d^2}\right)^{1/3}.
\label{Ti}
\end{equation}
Here, we point that while the axion couples to an enormous number of
fermions, $N_F \sim 10^{60}$, the effective coupling that governs the
axion decay remains perturbative.  To see this, note that the width of
the axion in the above example is roughly given by the Hubble constant
at $T\sim T_d\sim 100$~keV and is of order $10^{-18}$~eV $\sim 10^{-9}
m_a$. Hence, $N_F g^2_\chi \sim 10^{-8}$ which suggests a weakly coupled
and narrow axion at the relevant $\ord{m_a}$ energies for our preceding
analysis.

Let us assume $m_a \sim 100\, m \sim 10^{-9}$~eV and the reference parameters above.   We then find $T_i\sim \text{few}\times 100$~MeV.  Here, we note that an axion could in principle start oscillating once Hubble friction is diminished to $H \sim m_a \sim 10^{-9}$~eV, corresponding to $T_{osc} \sim 1$~GeV.  Since we have found that $T_i \lsim T_{osc}$ for a viable DM scenario, we interpret $T_i$ as the {\it reheat temperature} after inflation.  That is, with the above parameters ($N_F\sim10^{60}$, $T_{\rm NR}\sim$ keV, and $T_d\sim100\;T_{\rm NR}$), if the Universe evolved from a much larger temperature, oscillations of the axion would commence at around $T_{osc} \sim 1$~GeV which would dilute its energy density more than required.

\bibliography{main}

\end{document}